\documentclass[sigconf]{acmart}

\AtBeginDocument{%
  }

\copyrightyear{2023}
\acmYear{2023}
\setcopyright{acmlicensed}\acmConference[CIKM '23]{Proceedings of the 32nd ACM International Conference on Information and Knowledge Management}{October 21--25, 2023}{Birmingham, United Kingdom}
\acmBooktitle{Proceedings of the 32nd ACM International Conference on Information and Knowledge Management (CIKM '23), October 21--25, 2023, Birmingham, United Kingdom}
\acmPrice{15.00}
\acmDOI{10.1145/3583780.3615242}
\acmISBN{979-8-4007-0124-5/23/10}
\begin{document}

\title{FiBiNet++: Reducing Model Size by Low Rank Feature Interaction Layer for CTR Prediction}

\author{Pengtao Zhang}
\affiliation{%
  \institution{Sina Weibo}
  \city{Beijing}
  \country{China}
}
\email{zpt1986@126.com}

\author{Zheng Zheng}
\affiliation{%
  \institution{Brandeis University}
 \city{Waltham}
  \country{United States}
}
\email{zhengzheng@brandeis.edu}

\author{Junlin Zhang}
\affiliation{%
  \institution{Sina Weibo}
  \city{Beijing}
  \country{China}
}
\email{junlin6@staff.weibo.com}

\renewcommand{\shortauthors}{Pengtao Zhang, Zheng Zheng, \& Junlin Zhang}

\begin{abstract}

Click-Through Rate (CTR) estimation has become one of the most fundamental tasks in many real-world applications and various deep models have been proposed. Some research has proved that FiBiNet is one of the best performance models and outperforms all other models on Avazu dataset. However, the large model size of FiBiNet hinders its wider application. In this paper, we propose a novel FiBiNet++ model to redesign FiBiNet's model structure, which greatly reduces model size while further improves its performance. One of the primary techniques involves our proposed "Low Rank Layer" focused on feature interaction, which serves as a crucial driver of achieving a superior compression ratio for models. Extensive experiments on three public datasets show that FiBiNet++  effectively reduces non-embedding model parameters of FiBiNet by 12x to 16x on three datasets. On the other hand, FiBiNet++ leads to significant performance improvements compared to state-of-the-art CTR methods, including FiBiNet. The source code is in \url{https://github.com/recommendation-algorithm/FiBiNet}.

\end{abstract}






\maketitle

\section{Introduction}

Click-through rate (CTR) prediction plays important role in personalized advertising and recommender systems\cite{web1,graepel2010web,he2014practical,koren2009matrix,deeplight}. In recent years, a series of deep CTR models have been proposed to resolve this problem such as 
Wide \& Deep Learning\cite{cheng2016wide}, DeepFM\cite{guo2017deepfm}, DCN\cite{wang2017deep}, xDeepFM \cite{lian2018xdeepfm}, AutoInt\cite{song2019autoint}, DCN v2\cite{WangSCJLHC21} and FiBiNet\cite{HuangZZ19}. 
\begin{figure}[h]
  \centering
  \includegraphics[width=0.9\linewidth]{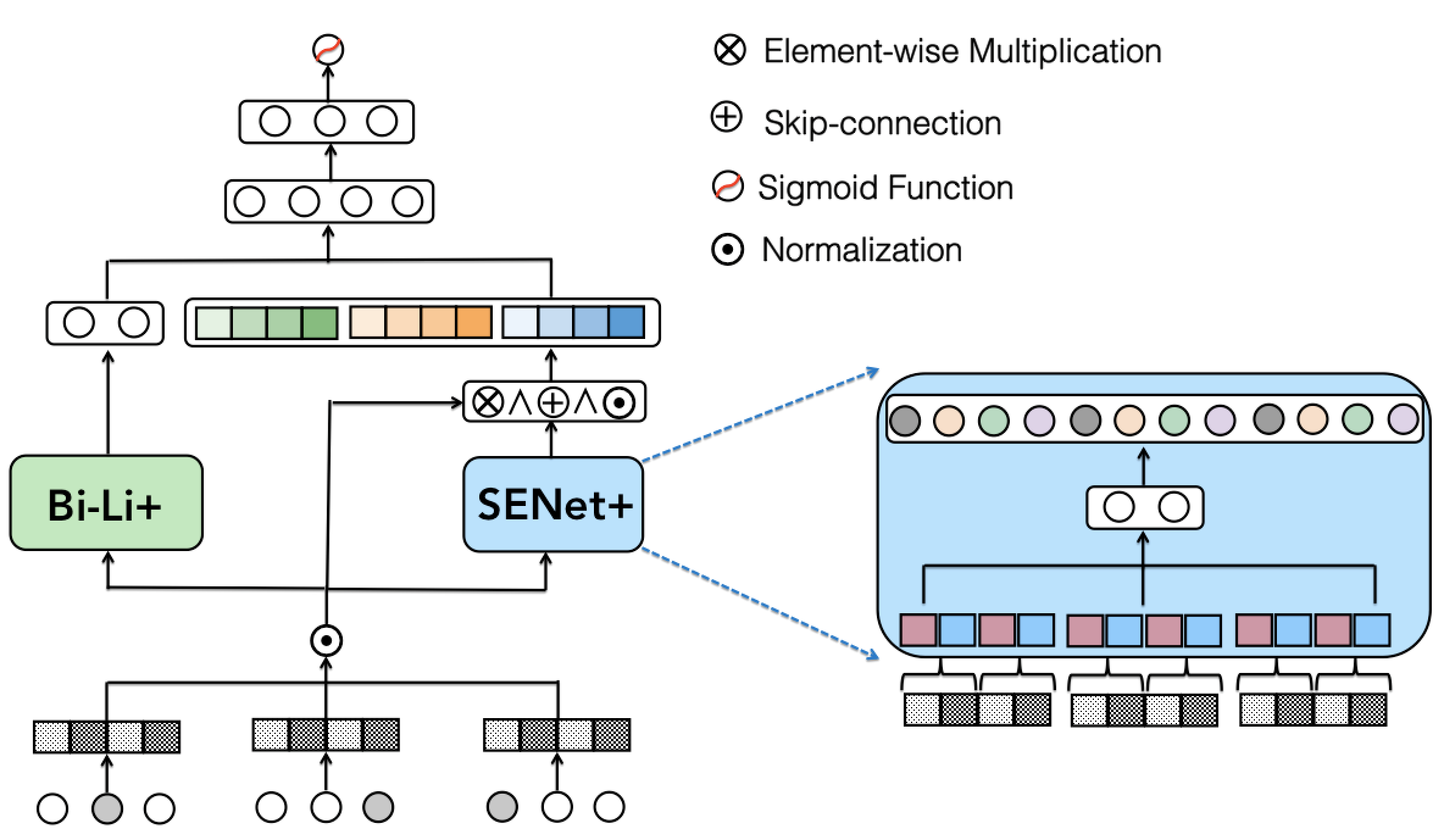}
  \caption{The Framework of FiBiNet++}
\end{figure}
Specifically, Wide \& Deep Learning\cite{cheng2016wide} jointly trains wide linear models and deep neural networks to combine the benefits of memorization and generalization for recommender systems. DeepFM\cite{guo2017deepfm} replaces the wide part of Wide \& Deep model with FM and shares the feature embedding between the FM and deep component. Some works explicitly introduce high-order feature interactions by sub-network. For example, Deep \& Cross Network (DCN)\cite{wang2017deep} and DCN v2\cite{WangSCJLHC21} efficiently capture feature interactions of bounded degrees in an explicit fashion. The eXtreme Deep Factorization Machine (xDeepFM) \cite{lian2018xdeepfm} also models the low-order and high-order feature interactions in an explicit way by proposing a novel Compressed Interaction Network (CIN) part. Similarly, 
FiBiNet\cite{HuangZZ19} dynamically learns the importance of features via the Squeeze-Excitation network (SENET) and feature interactions via bi-linear function.

Though many models have been proposed, seldom works fairly compare these models' performance. FuxiCTR\cite{jieming-2009-05794} performs open benchmarking for CTR prediction and experimental results\cite{jieming-2009-05794} show that FiBiNet is one of the best performance models, which outperforms all other 23 models on Avazu dataset.

However, we argue that FiBiNet has too many model parameters, which hinders its wider usage in real-life applications.  In real-world systems, both the smaller size and the cost of training and inference times are important factors to be considered. Therefore, our works aims to redesign the model structure to greatly reduce model size while improving its performance. 

In this paper, we propose a novel FiBiNet++ model to address these challenges as shown in Figure 1. First, we reconstruct the model structure by removing the bi-linear module on SENet and the linear part in FiBiNet, which directly reduces model parameters. More importantly, we upgrade the bi-linear function into the bi-linear+ module by changing the hadamard product to inner product and bringing a "Low Rank Layer" on feature interaction into it. In Section 3.2, we will demonstrate that the proposed "Low Rank Layer" is primarily responsible for high model compression ratio. Our inspiration for this approach stems from the LoRA\cite{lora}, which reveals that LLM models possess a low rank "intrinsic dimension," enabling them to learn effectively even after undergoing random projection into a smaller subspace. We put forth the hypothesis that feature interactions in CTR models similarly exhibit a low intrinsic rank during training and propose incorporating a "Low Rank Feature Interaction Layer" into bi-linear+ modules, which greatly reduces model parameters while keeps model's performance. Finally, we introduce feature normalization and the upgraded SENet+ module to further enhance model performance.


We summarize our major contributions as below:
(1) The proposed FiBiNet++ greatly reduces model size of FiBiNet by 12x to 16x on three datasets. 
(2) Compared with FiBiNet, our proposed FiBiNet++ model increases mode's training and reference efficiencies by +37.50\% to 81.03\% on three datasets.
(3) FiBiNet++ yields remarkable improvements compared to state-of-the-art models.

\section{Preliminaries}
DNN model is always used as a sub-component in most current DNN ranking systems\cite{xiao2017attentional,guo2017deepfm,lian2018xdeepfm,wang2017deep,HuangZZ19,covington2016deep,qu2016} and it contains two components: feature embedding and MLP.

\textbf{(1) Feature Embeddding.}
We map one-hot sparse features to dense, low-dimensional embedding vectors and obtain feature embedding $\mathbf{v}_i$ for one-hot vector $\mathbf{x}_i$ via:
$\mathbf{v}_i = \mathbf{W}_e\mathbf{x}_i\in\mathbb{R}^{1\times d}$
, where $\mathbf{W}_e \in\mathbb{R}^{n\times d}$ is the embedding matrix of $n$ features and $d$ is the dimension of field embedding.

\textbf{(2) MLP.}
To learn high-order feature interactions, multiple feed-forward layers are stacked on the concatenation of dense features represented as  $\mathbf{H}_0 = concat[\mathbf{v}_{1},\mathbf{v}_{2},...,\mathbf{v}_{f}]$, where $f$ denotes field number. Then, the feed forward process of MLP is 
$ \mathbf{H}_l = ReLU(\mathbf{W}_l\mathbf{H}_{l-1} + \beta_l)$ , where $l$ is the depth and ReLU is the activation function. $\mathbf{W}_l, \beta_l, \mathbf{H}_l$ are weighting matrix, bias and output of the $l$-th layer. 

For binary classifications, the loss function of CTR prediction is the log loss:
\begin{equation}
  \mathcal{L} = -\frac{1}{N}\sum^N_{i=1}y_i\log(\hat{y}_i)+(1-y_i)\log(1-\hat{y}_i) 
\end{equation}
where $N$ is the total number of training instances, $y_i$ is the ground truth of $i$-th instance and $\hat{y}_i$ is the predicted CTR.  

\section{Our Proposed Model}
The architecture of the proposed FiBiNet++ is shown in Figure 1. The original feature embedding is first normalized before being sent to the following components. Then, Bi-linear+ module models feature interactions and SENet+ module computes bit-wise feature importance. The outputs of two branches are concatenated as input of the following MLP layers.
\subsection{FiBiNet++}

\textbf{Feature Normalization.}
We introduce feature normalization into FiBiNet++ to enhance model's training stability as follows:
\begin{equation}
  \mathbf{N}(\mathbf{V}) = concat[\mathbf{N}(\mathbf{v}_{1}),\mathbf{N}(\mathbf{v}_{2}),...,\mathbf{N}(\mathbf{v}_{f})]\in \mathbb{R}^{1\times fd}
\end{equation}
where  $\mathbf{N}\left(\cdot\right)$ is layer normalization\cite{ln2016} for numerical feature and batch normalization\cite{bn2015} operation for categorical feature.

\textbf{Bi-Linear+ Module.}
FiBiNet models interaction between feature $\mathbf{x}_i$ and feature $\mathbf{x}_j$ by bi-linear function which introduces an extra learned matrix $\mathbf{W}$ as follows:
\begin{equation}
\mathbf{p}_{i,j} = \mathbf{v}_{i}\circ\mathbf{W}\otimes\mathbf{v}_{j} \in \mathbb{R}^{1\times d}
\end{equation}
where $\circ$ and $\otimes$ denote inner product and element-wise hadamard product, respectively. In order to effectively reduce model size, we upgrade bi-linear function into bi-linear+ module by following two methods. First, the hadamard product is replaced by another inner product as: $\mathbf{p}_{i,j} = \mathbf{v}_{i}\circ\mathbf{W}\circ\mathbf{v}_{j} \in \mathbb{R}^{1\times 1}$
. We think the feature interaction is rather sparse and one bit information as representation is enough instead of a vector. It's easy to see the parameters of $\mathbf{p}_{i,j}$ decrease greatly from  $\mathbf{d}$ dimensional vector to $1$ bit for each feature interaction. Suppose the input instance has $f$ fields and we have the following  vector after bi-linear feature interaction:
\begin{equation}
         \mathbf{P} = concat[\mathbf{p}_{1,2},\mathbf{p}_{1,3},......,\mathbf{p}_{f-1,f}] \in \mathbb{R}^{1\times \frac{f\times (f-1)}{2} }    
 \end{equation}

Inspired by LoRA\cite{lora}, which has demonstrated that LLM models possess a low "intrinsic dimension" and exhibit efficient learning despite undergoing random projections to smaller subspaces, we posit that updates to the feature interaction layer during training also exhibit a low "intrinsic rank." Therefore, we propose integrating a thin "Low Rank Layer" into Bi-Linear+, thereby reducing model parameters significantly while maintaining optimal model performance. Specifically, we introduce "Low Rank Layer" stacking on feature interaction vector  $\mathbf{P}$ as follows:
 \begin{equation}
\mathbf{H}^{LRL} = \sigma_1\left(\mathbf{W}_{1}\mathbf{P}\right )\in \mathbb{R}^{1\times m}
\end{equation}
where $\mathbf{W}_1 \in \mathbb{R}^{ m\times \frac{f\times (f-1)}{2} } $ is a learning matrix of thin MLP layer with small size  $m$ and $\sigma_1\left(\cdot\right) $ is an identity function. "Low Rank Layer" projects feature interactions from  sparse space into low rank space to greatly reduce the storage.

\textbf{SENet+ Module.}
SENet+ module consists of four phases: squeeze, excitation, re-weight and fuse.
\textbf{(1) Squeeze.} SENet collects one bit information by mean pooling from each feature embedding as  "summary statistics". However, we improve the original squeeze step by providing more useful information. Specifically, we first segment each normalized feature embedding $\mathbf{v}_{i} \in\mathbb{R}^{1\times d}$ into $\mathbf g $  groups, which is a hyper-parameter, as:
 $\mathbf{v}_i = concat[\mathbf{v}_{i,1},\mathbf{v}_{i,2},......,\mathbf{v}_{i,g}] $
, where $\mathbf{v}_{i,j}\in \mathbb{R}^{1\times \frac{d}{g}}$ means information in the j-th group of the i-th feature. Let $\mathbf{k}=\frac{d}{g} $ denotes the size of each group. Then, we select the max value  $\mathbf{z}_{i,j}^{max}$and average pooling value $\mathbf{z}_{i,j}^{avg}$ in $\mathbf{v}_{i,j}$ as representation of the group as:
$\mathbf{z}_{i,j}^{max} = \max \limits_{t} \left\{ \mathbf{v}_{i,j}^t \right\}_{t=1}^{k}$ and $
          \mathbf{z}_{i,j}^{avg} =\frac{1}{k}\sum_{t=1}^k \mathbf{v}_{i,j}^t$
. The concatenated representative information of each group forms the "summary statistic" $\mathbf{Z}_{i} $ of feature embedding  $\mathbf{v}_{i} $:
\begin{equation}
         \mathbf{Z}_{i} = concat[\mathbf{z}_{i,1}^{max},\mathbf{z}_{i,1}^{avg},\mathbf{z}_{i,2}^{max},\mathbf{z}_{i,2}^{avg},......,\mathbf{z}_{i,g}^{max},\mathbf{z}_{i,g}^{avg}] \in \mathbb{R}^{1\times 2g}
 \end{equation}
Finally, we can concatenate each feature's summary statistic 
$\mathbf{Z} = concat[\mathbf{Z}_{1},\mathbf{Z}_{2},......,\mathbf{Z}_{f}] \in \mathbb{R}^{1\times 2gf}$
as the input of SENet+ module.

\textbf{(2) Excitation.} The excitation step in SENet computes each feature's weight according to the statistic vector $\mathbf{Z} $, which is a field-wise attention.  However, we improve this step by changing the field-wise attention into a more fine-grained bit-wise attention. Similarly, we also use two fully connected (FC) layers to learn the weights as follows:
$\mathbf{A} = \sigma_3\left(\mathbf{W}_{3}\sigma_2\left(\mathbf{W}_{2}\mathbf{Z}\right )\right)\in \mathbb{R}^{1\times fd}$
, where  $\mathbf{W}_2 \in \mathbb{R}^{ \frac{2gf}{r}\times 2gf} $ denotes 
learning parameters of the first FC layer, which is a thin layer and $r$ is reduction ratio. $\mathbf{W}_3 \in \mathbb{R}^{ fd \times\frac{2gf}{r}} $ means learning parameters of the second FC layer, which is a wider layer with a size of $fd$. Here $\sigma_2\left(\cdot\right) $ is  $ReLu \left(\cdot\right)$ and $\sigma_3\left(\cdot\right) $ is an identity function without non-linear transformation.In this way, each bit in input embedding can dynamically learn the corresponding attention score provided by $\mathbf{A}$. 

\textbf{(3) Re-Weight.} Re-weight step does element-wise multiplication between the original field embedding and the learned attention scores as follows:
$\mathbf{V}^w = \mathbf{A}\otimes\mathbf{N}(\mathbf{V})\in \mathbb{R}^{1\times fd}$
, where  $\otimes$ is an element-wise multiplication between two vectors and $\mathbf{N}(\mathbf{V})$ denotes original embedding after normalization.

\textbf{(4) Fuse.} An extra "fuse" step is introduced in order to better fuse the information contained both in original feature embedding and weighted embedding. Specifically, we first use skip-connection to merge two embedding as follows:
$\mathbf{v}_i^s = \mathbf{v}_i^o\oplus\mathbf{v}_i^w$
, where $\mathbf{v}_i^o$ donates the i-th normalized feature embedding, $\mathbf{v}_i^w$ denotes embedding after re-weight step, $\oplus$ is an element-wise addition operation. Then, another feature normalization is applied on feature embedding $\mathbf{v}_i^s$ for a better representation:
$\mathbf{v}_i^u = \mathbf{LN}(\mathbf{v}_i^s)$
. Note we adopt layer normalization here no matter what type of feature it belongs to, numerical  or categorical feature. Finally, we concatenate all the fused embeddings as the output of the SENet+ module:
      \begin{equation}
         \mathbf{V}^{SENet+}= concat[\mathbf{v}_1^u,\mathbf{v}_2^u,...,\mathbf{v}_f^u]\in \mathbb{R}^{1\times fd}
    \end{equation}

\textbf{Concatenation Layer.}
We concatenate the output of two branches to form the input of the following MLP layers:
\begin{equation}
\mathbf{H}_0 = concat[\mathbf{H}^{LRL},\mathbf{V}^{SENet+}]
\end{equation}

\subsection{Discussion}
In this section, we discuss the model size difference  between FiBiNet and FiBiNet++. Note only non-embedding parameter is considered, which really demonstrates model complexity.

The major parameter of FiBiNet comes from two components: one is the connection between the first MLP layer and the output of two bi-linear modules, and the other is the linear part. Suppose we denote $h=400$ as the size of the first MLP layer, $f=50$ as field number, $d=10$ as feature embedding size, and  $t=1 \ million$ as feature number. Therefore, the parameter number in these two parts is nearly 10.8 million:
\begin{equation}
  \mathbf{T}^{FiBiNet} = \begin{matrix} \underbrace{ \mathbf{f}\times (\mathbf{f-1}) \times \mathbf{d} \times \mathbf{h} }_{MLP \ and \ bi-linear}+ \underbrace{ t }_{ linear  } \end{matrix}=10.8 \ millions
\end{equation}
For FiBiNet++, the majority of model parameter comes from following three components: the connection between the first MLP layer and embedding produced by SENet+ module(1-th part), the connection between the first MLP layer and "Low Rank Layer"(2-th part), and parameters between "Low Rank Layer" and bi-linear feature interaction results(3-th part). Let $m=50$ denote the size of "Low Rank Layer". We have the parameter number of these components as follows:
\begin{equation}
  \mathbf{T}^{FiBiNet++} = \begin{matrix} \underbrace{ \mathbf{f}\times  \mathbf{d} \times \mathbf{h} }_{1-th \ part}+ \underbrace{\mathbf{m} \times \mathbf{h}  }_{ 2-th \ part  } + \underbrace{\mathbf{\frac{\mathbf{f}\times (\mathbf{f-1}) }{2} }\times \mathbf{m}  }_{ 3-th \ part }\end{matrix}=0.28 \ millions
\end{equation}
We can see that the above-mentioned methods to reduce model size greatly decrease model size from 10.8 million to 0.28 million, which is nearly 39x model compression. In addition, the larger the field number $f$ is, the larger the model compression ratio we can achieve. It's easy to see that "Low Rank Layer" is the key to the high compression ratio  while it can also be applied into other CTR models with long feature interaction layers such as ONN\cite{onn} and FAT-DeepFFM\cite{fat} for feature interaction compression.

\section{Experimental Results}
\subsection{Experiment Setup}
\textbf{Datasets}
Three datasets are used in our experiments and we randomly split instances by 8:1:1 for training, validation and testing:
\textbf{(1) Criteo\footnote{Criteo \url{http://labs.criteo.com/downloads/download-terabyte-click-logs/}}:}
  As a display ad dataset, there are 26 anonymous categorical fields and 13 continuous feature fields.
\textbf{(2) Avazu\footnote{Avazu \url{http://www.kaggle.com/c/avazu-ctr-prediction}}:}
The Avazu dataset contains 23 fields that indicate elements of a single ad impression.
\textbf{(3) KDD12\footnote{KDD12 \url{https://www.kaggle.com/c/kddcup2012-track2}}:}
  KDD12 dataset has 13 fields spanning from user id to ad position for a clicked data.
  
\begin{table}
\centering
\caption{Overall performance (AUC) of different models}
\resizebox{85mm}{23mm}{
\begin{tabular}{l|ccccccc}
\hline

  & \multicolumn{2}{c}{\textbf{Avazu}}  & \multicolumn{2}{c}{\textbf{Criteo} } & \multicolumn{2}{c}{\textbf{KDD12}} \\
\hline
Model & AUC(\%)  &  Paras. & AUC(\%) & Paras.  & AUC(\%) & Paras.   \\
\hline
FM       & 78.17 &1.54M   & 78.97 &1.0M & 77.65  & 5.46M  \\
DNN       & 78.67 &\underline{0.74M}   & 80.73 &\underline{0.48M} & 79.54  & \underline{0.37M}  \\
DeepFM       & 78.64 &2.29M   & 80.58 &1.48M & 79.40  & 5.84M  \\
xDeepFM       & 78.88 &4.06M   & 80.64 &4.90M & 79.51  & 6.91M  \\
DCN       & 78.68 &0.75M   & 80.73 &0.48M & 79.58  & 0.37M  \\
AutoInt+       & 78.62 &0.77M   & 80.78 &0.48M & \underline{79.69}  & 0.38M  \\
DCN v2       & 78.98 &4.05M   & \underline{80.88} &0.65M & 79.66  & 0.39M  \\
\hline
FiBiNet       & \underline{79.12} &10.27M   & 80.73 &7.25M & 79.52  & 6.41M  \\

  FiBiNet++ & \textbf{79.15} &0.81M   & \textbf{81.10} &0.56M & \textbf{79.98}  & 0.40M   \\
   Improv. & +0.03 &12.7x   & +0.37 &12.9x & +0.46  & 16x   \\
\hline
\end{tabular}}
\label{tab:overalperformance}
\end{table}
\begin{figure*}[h]
  \centering
  \includegraphics[width=0.8\linewidth]{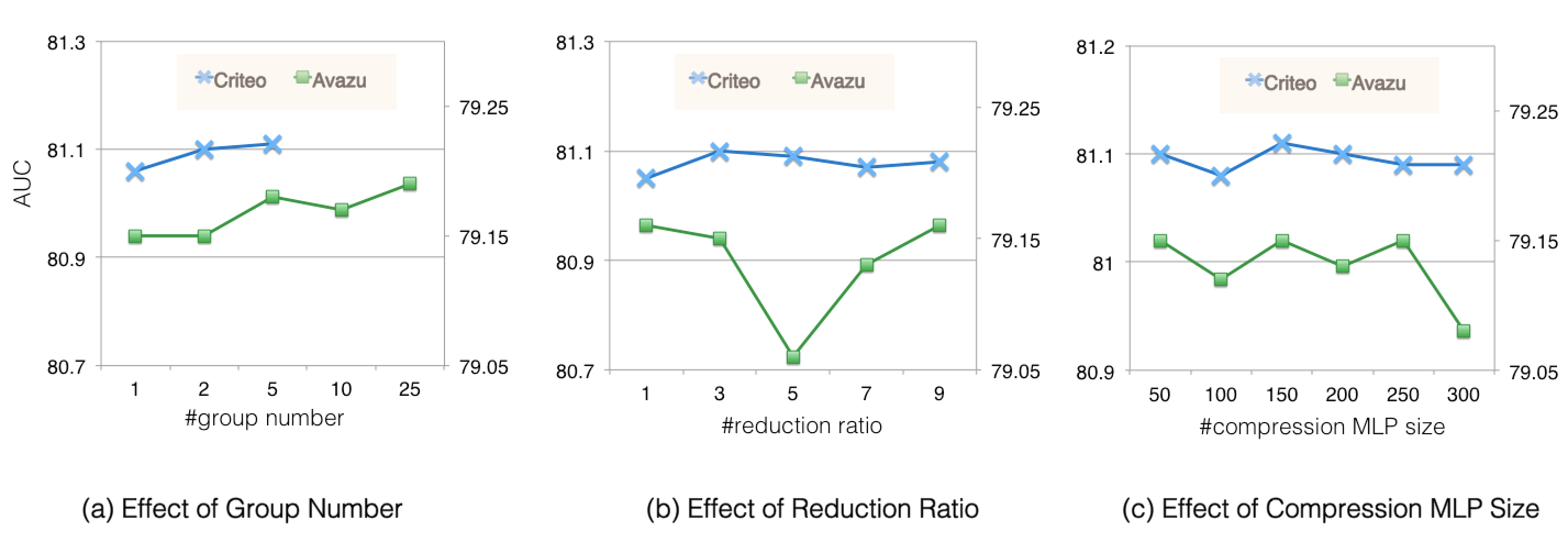}
  \caption{Effect of Hyper-Parameters}
\end{figure*}

\textbf{Models for Comparisons}
We compare the performance of the FM\cite{rendle2010factorization}, DNN, DeepFM\cite{guo2017deepfm}, DCN \cite{wang2017deep}, AutoInt \cite{song2019autoint}, DCN V2 \cite{WangSCJLHC21}, xDeepFM \cite{lian2018xdeepfm} and FiBiNet \cite{HuangZZ19} models as baselines and AUC is used as the evaluation metric.
\begin{figure}[h]
  \centering
  \includegraphics[width=0.8\linewidth]{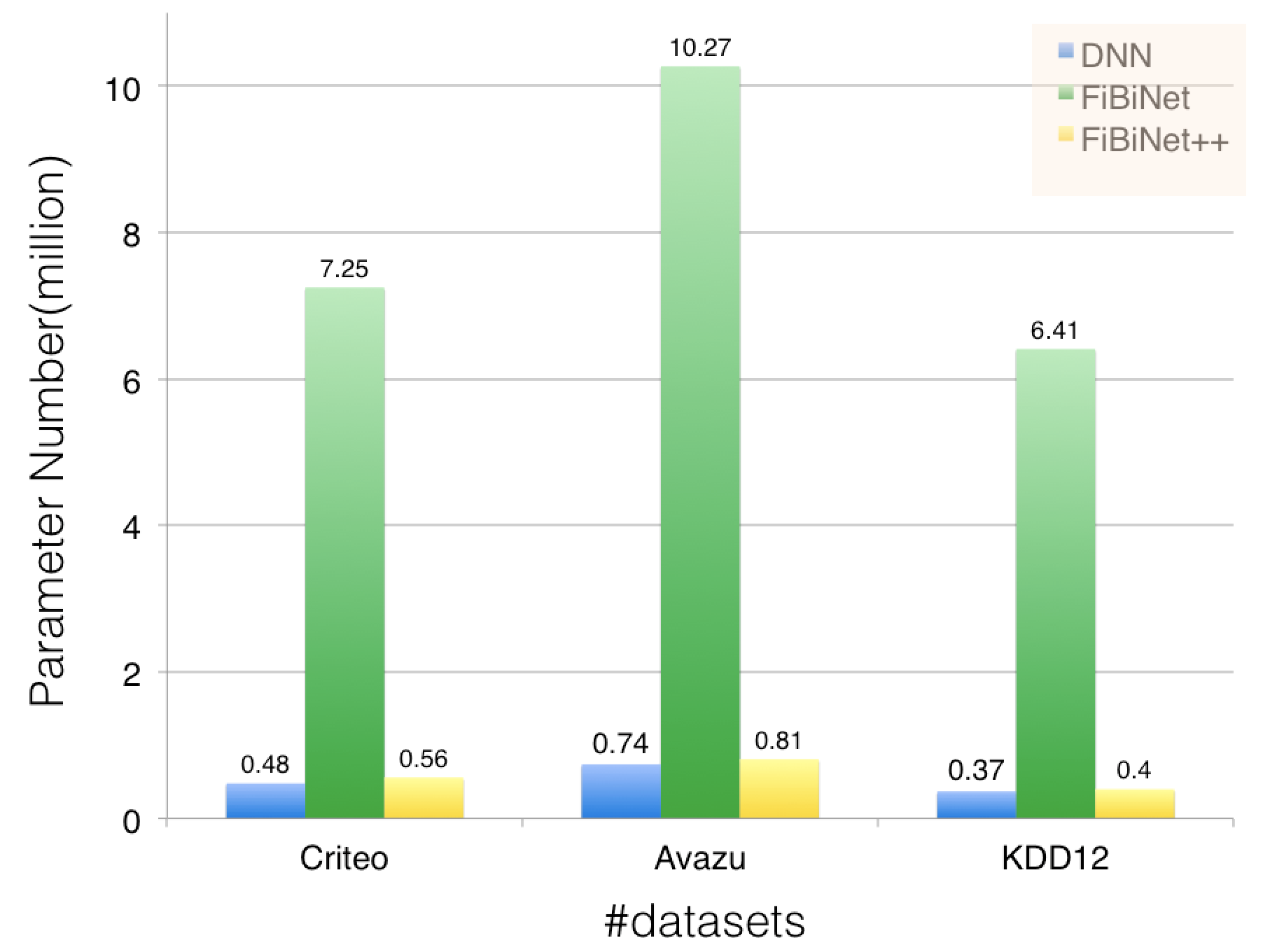}
  \caption{Comparison of Model Size}
\end{figure}

\textbf{Implementation Details}
For the optimization method, we use the Adam with a mini-batch size of $1024$ and $0.0001$ as learning rates. We make the dimension of field embedding for all models to be a fixed value of $10$ for Criteo dataset,  $50$ for Avazu dataset and $10$ for KDD12 dataset.  For models with DNN part, the depth of hidden layers is set to $3$, the number of neurons per layer is $400$, and all activation functions are ReLU. In SENet+, the reduction ratio is set to 3 and the group number is 2 as the default settings. In Bi-linear+ module, we set size of the low rank layer as 50. 

\subsection{Results and Analysis}
\textbf{Performance Comparison.}  Table \ref{tab:overalperformance} shows the performances of different SOTA baselines and FiBiNet++.  The best results are in bold, and the best baseline results are underlined. We can see that:(1) FiBiNET++ model outperforms all the compared SOTA methods and achieves the best performance on all three benchmarks. (2) Among all the strong baselines with a similar amount of parameters such as Autoint+, DCN v2 and DeepFM, FiBiNet++ is the best performance model on all three datasets. This demonstrates the reason why FiBiNet++ outperforms other models is because of its designed components instead of more parameter. (3) Compared with FiBiNet model, FiBiNet++ can achieve better performance on all datasets though it has much fewer parameters, which indicates that our proposed methods to enhance model performance are effective.

\textbf{Model Size Comparison.} We compare the non-embedding model size of different methods in Table 1 and Figure 3. FiBiNet++ provides orders of magnitude improvement in model size while improving the quality of the model compared with FiBiNet. Specifically, FiBiNet++ reduce the model size of FiBiNet by 12.7x, 12.9x and 16x in terms of the number of parameters on three datasets, respectively, which demonstrates that our proposed methods to reduce model parameter in this paper are effective. Now FiBiNet++ has a comparable model size with DNN model while outperforms all other models on three datasets at the same time. 

\textbf{Training/Reference Efficiency.} 
Efficiency is an essential concern in industrial applications and we conduct experiments to compare the training and inference time between our proposed FiBiNet++ and FiBiNet. We leverage time(millisecond) of processing one batch of examples during the training and reference as an efficiency metric. The average training and  inference times of the two models are illustrated in Table 2. Compared with FiBiNet model, the training efficiency of FiBiNet++ increases by 58.76\%, 62.30\% and 39.39\% while reference efficiency increases by 41.66\%, 81.03\% and 37.50\% on three datasets respectively. Our FiBiNet++ model shows a significant advantage in training and inference efficiency, which makes it more practical to be applied in real life.

\textbf{Hyper-Parameters of FiBiNet++.} Next, we study hyperparameter sensitivity of FiBiNet++. 
\textbf{ (1) Group Number.}  Figure 2a shows a slight performance increase with the increase of group number, which indicates that more group number benefits model performance because we can input more useful information in feature embedding into SENet+ module.

\textbf{ (2) Reduction Ratio.} We conduct some experiments to adjust the reduction ratio in SENet+ module from $1$ to $9$ and Figure 2b shows the result. It can be seen that the performance is better if we set the reduction ratio  to 3 or 9.
\textbf{ (3) Size of Low Rank Layer.} The results in Figure 2c show the impact when we adjust the size of the Low Rank Layer in bi-linear+ module. We can observe that the performance begins to decrease when the size is set greater than 150, which demonstrates the feature interaction represented in low rank space indeed works.
\begin{table}
\centering
\caption{Training and reference efficiency comparison}
\resizebox{80mm}{!}{
\begin{tabular}{l|ccccccc}
\hline

  & \multicolumn{2}{c}{\textbf{Avazu}}  & \multicolumn{2}{c}{\textbf{Criteo} } & \multicolumn{2}{c}{\textbf{KDD12}} \\
\hline
Model &Train(ms)  &  Refer(ms) & Train(ms)  &  Refer(ms)  & Train(ms)  &  Refer(ms)  \\
\hline
FiBiNet       & 97 &12   & 191 &58 & 33  & 8  \\

  FiBiNet++ & 40 &7   & 72 &11 & 20 & 5   \\
  \hline
   Improv. & +58.76\% &+41.66\%   & +62.30\% &+81.03\% & +39.39\%  & +37.50\%   \\
\hline
\end{tabular}}
\label{tab:training}
\end{table}

\section{Conclusion}
In this paper, we propose FiBiNet++ model in order to greatly reduce the model size while improving the model performance. Experimental results show that FiBiNet++ provides orders of magnitude improvement in model size while improving the quality of the model.

\bibliographystyle{ACM-Reference-Format}
\balance
\bibliography{sample-base}


\begin{thebibliography}{22}


\ifx \showCODEN    \undefined \def \showCODEN     #1{\unskip}     \fi
\ifx \showDOI      \undefined \def \showDOI       #1{#1}\fi
\ifx \showISBNx    \undefined \def \showISBNx     #1{\unskip}     \fi
\ifx \showISBNxiii \undefined \def \showISBNxiii  #1{\unskip}     \fi
\ifx \showISSN     \undefined \def \showISSN      #1{\unskip}     \fi
\ifx \showLCCN     \undefined \def \showLCCN      #1{\unskip}     \fi
\ifx \shownote     \undefined \def \shownote      #1{#1}          \fi
\ifx \showarticletitle \undefined \def \showarticletitle #1{#1}   \fi
\ifx \showURL      \undefined \def \showURL       {\relax}        \fi
\providecommand\bibfield[2]{#2}
\providecommand\bibinfo[2]{#2}
\providecommand\natexlab[1]{#1}
\providecommand\showeprint[2][]{arXiv:#2}

\bibitem[Ba et~al\mbox{.}(2016)]%
        {ln2016}
\bibfield{author}{\bibinfo{person}{Jimmy~Lei Ba}, \bibinfo{person}{Jamie~Ryan
  Kiros}, {and} \bibinfo{person}{Geoffrey~E Hinton}.}
  \bibinfo{year}{2016}\natexlab{}.
\newblock \showarticletitle{Layer normalization}.
\newblock \bibinfo{journal}{\emph{arXiv preprint arXiv:1607.06450}}
  (\bibinfo{year}{2016}).
\newblock


\bibitem[Cheng et~al\mbox{.}(2016)]%
        {cheng2016wide}
\bibfield{author}{\bibinfo{person}{Heng-Tze Cheng}, \bibinfo{person}{Levent
  Koc}, \bibinfo{person}{Jeremiah Harmsen}, \bibinfo{person}{Tal Shaked},
  \bibinfo{person}{Tushar Chandra}, \bibinfo{person}{Hrishi Aradhye},
  \bibinfo{person}{Glen Anderson}, \bibinfo{person}{Greg Corrado},
  \bibinfo{person}{Wei Chai}, \bibinfo{person}{Mustafa Ispir}, {et~al\mbox{.}}}
  \bibinfo{year}{2016}\natexlab{}.
\newblock \showarticletitle{Wide \& deep learning for recommender systems}. In
  \bibinfo{booktitle}{\emph{Proceedings of the 1st workshop on deep learning
  for recommender systems}}. ACM, \bibinfo{pages}{7--10}.
\newblock


\bibitem[Covington et~al\mbox{.}(2016)]%
        {covington2016deep}
\bibfield{author}{\bibinfo{person}{Paul Covington}, \bibinfo{person}{Jay
  Adams}, {and} \bibinfo{person}{Emre Sargin}.}
  \bibinfo{year}{2016}\natexlab{}.
\newblock \showarticletitle{Deep neural networks for youtube recommendations}.
  In \bibinfo{booktitle}{\emph{Proceedings of the 10th ACM conference on
  recommender systems}}. ACM, \bibinfo{pages}{191--198}.
\newblock


\bibitem[Deng et~al\mbox{.}(2021)]%
        {deeplight}
\bibfield{author}{\bibinfo{person}{Wei Deng}, \bibinfo{person}{Junwei Pan},
  \bibinfo{person}{Tian Zhou}, \bibinfo{person}{Deguang Kong},
  \bibinfo{person}{Aaron Flores}, {and} \bibinfo{person}{Guang Lin}.}
  \bibinfo{year}{2021}\natexlab{}.
\newblock \showarticletitle{Deeplight: Deep lightweight feature interactions
  for accelerating ctr predictions in ad serving}. In
  \bibinfo{booktitle}{\emph{Proceedings of the 14th ACM international
  conference on Web search and data mining}}. \bibinfo{pages}{922–--930}.
\newblock


\bibitem[Graepel et~al\mbox{.}(2010)]%
        {graepel2010web}
\bibfield{author}{\bibinfo{person}{Thore Graepel},
  \bibinfo{person}{Joaquin~Quinonero Candela}, \bibinfo{person}{Thomas
  Borchert}, {and} \bibinfo{person}{Ralf Herbrich}.}
  \bibinfo{year}{2010}\natexlab{}.
\newblock \showarticletitle{Web-scale bayesian click-through rate prediction
  for sponsored search advertising in microsoft's bing search engine}.
  Omnipress.
\newblock


\bibitem[Guo et~al\mbox{.}(2017)]%
        {guo2017deepfm}
\bibfield{author}{\bibinfo{person}{Huifeng Guo}, \bibinfo{person}{Ruiming
  Tang}, \bibinfo{person}{Yunming Ye}, \bibinfo{person}{Zhenguo Li}, {and}
  \bibinfo{person}{Xiuqiang He}.} \bibinfo{year}{2017}\natexlab{}.
\newblock \showarticletitle{DeepFM: a factorization-machine based neural
  network for CTR prediction}.
\newblock \bibinfo{journal}{\emph{arXiv preprint arXiv:1703.04247}}
  (\bibinfo{year}{2017}).
\newblock


\bibitem[He et~al\mbox{.}(2014)]%
        {he2014practical}
\bibfield{author}{\bibinfo{person}{Xinran He}, \bibinfo{person}{Junfeng Pan},
  \bibinfo{person}{Ou Jin}, \bibinfo{person}{Tianbing Xu}, \bibinfo{person}{Bo
  Liu}, \bibinfo{person}{Tao Xu}, \bibinfo{person}{Yanxin Shi},
  \bibinfo{person}{Antoine Atallah}, \bibinfo{person}{Ralf Herbrich},
  \bibinfo{person}{Stuart Bowers}, {et~al\mbox{.}}}
  \bibinfo{year}{2014}\natexlab{}.
\newblock \showarticletitle{Practical lessons from predicting clicks on ads at
  facebook}. In \bibinfo{booktitle}{\emph{Proceedings of the Eighth
  International Workshop on Data Mining for Online Advertising}}. ACM,
  \bibinfo{pages}{1--9}.
\newblock


\bibitem[Hu et~al\mbox{.}(2021)]%
        {lora}
\bibfield{author}{\bibinfo{person}{Edward~J Hu}, \bibinfo{person}{Yelong Shen},
  \bibinfo{person}{Phillip Wallis}, \bibinfo{person}{Zeyuan Allen-Zhu},
  \bibinfo{person}{Yuanzhi Li}, \bibinfo{person}{Shean Wang},
  \bibinfo{person}{Lu Wang}, {and} \bibinfo{person}{Weizhu Chen}.}
  \bibinfo{year}{2021}\natexlab{}.
\newblock \showarticletitle{Lora: Low-rank adaptation of large language
  models}.
\newblock \bibinfo{journal}{\emph{arXiv preprint arXiv:2106.09685}}
  (\bibinfo{year}{2021}).
\newblock


\bibitem[Huang et~al\mbox{.}(2019)]%
        {HuangZZ19}
\bibfield{author}{\bibinfo{person}{Tongwen Huang}, \bibinfo{person}{Zhiqi
  Zhang}, {and} \bibinfo{person}{Junlin Zhang}.}
  \bibinfo{year}{2019}\natexlab{}.
\newblock \showarticletitle{FiBiNET: combining feature importance and bilinear
  feature interaction for click-through rate prediction}. In
  \bibinfo{booktitle}{\emph{Proceedings of the 13th {ACM} Conference on
  Recommender Systems, RecSys 2019, Copenhagen, Denmark, September 16-20,
  2019}}. \bibinfo{publisher}{{ACM}}, \bibinfo{pages}{169--177}.
\newblock
\urldef\tempurl%
\url{https://doi.org/10.1145/3298689.3347043}
\showURL{%
\tempurl}


\bibitem[Ioffe and Szegedy(2015)]%
        {bn2015}
\bibfield{author}{\bibinfo{person}{Sergey Ioffe} {and}
  \bibinfo{person}{Christian Szegedy}.} \bibinfo{year}{2015}\natexlab{}.
\newblock \showarticletitle{Batch normalization: Accelerating deep network
  training by reducing internal covariate shift}.
\newblock \bibinfo{journal}{\emph{arXiv preprint arXiv:1502.03167}}
  (\bibinfo{year}{2015}).
\newblock


\bibitem[Koren et~al\mbox{.}(2009)]%
        {koren2009matrix}
\bibfield{author}{\bibinfo{person}{Yehuda Koren}, \bibinfo{person}{Robert
  Bell}, {and} \bibinfo{person}{Chris Volinsky}.}
  \bibinfo{year}{2009}\natexlab{}.
\newblock \showarticletitle{Matrix factorization techniques for recommender
  systems}.
\newblock \bibinfo{journal}{\emph{Computer}} \bibinfo{number}{8}
  (\bibinfo{year}{2009}), \bibinfo{pages}{30--37}.
\newblock


\bibitem[Lian et~al\mbox{.}(2018)]%
        {lian2018xdeepfm}
\bibfield{author}{\bibinfo{person}{Jianxun Lian}, \bibinfo{person}{Xiaohuan
  Zhou}, \bibinfo{person}{Fuzheng Zhang}, \bibinfo{person}{Zhongxia Chen},
  \bibinfo{person}{Xing Xie}, {and} \bibinfo{person}{Guangzhong Sun}.}
  \bibinfo{year}{2018}\natexlab{}.
\newblock \showarticletitle{xdeepfm: Combining explicit and implicit feature
  interactions for recommender systems}. In
  \bibinfo{booktitle}{\emph{Proceedings of the 24th ACM SIGKDD International
  Conference on Knowledge Discovery \& Data Mining}}. ACM,
  \bibinfo{pages}{1754--1763}.
\newblock


\bibitem[McMahan et~al\mbox{.}(2013)]%
        {web1}
\bibfield{author}{\bibinfo{person}{H.~Brendan McMahan}, \bibinfo{person}{Gary
  Holt}, \bibinfo{person}{D. Sculley}, \bibinfo{person}{Michael Young},
  \bibinfo{person}{Dietmar Ebner}, \bibinfo{person}{Julian Grady},
  \bibinfo{person}{Lan Nie}, \bibinfo{person}{Todd Phillips},
  \bibinfo{person}{Eugene Davydov}, \bibinfo{person}{Daniel Golovin}, {and}
  \bibinfo{person}{et al.}} \bibinfo{year}{2013}\natexlab{}.
\newblock \showarticletitle{Ad Click Prediction: A View from the Trenches}. In
  \bibinfo{booktitle}{\emph{Proceedings of the 19th ACM SIGKDD International
  Conference on Knowledge Discovery and Data Mining}} (Chicago, Illinois, USA)
  \emph{(\bibinfo{series}{KDD ’13})}. \bibinfo{publisher}{Association for
  Computing Machinery}, \bibinfo{address}{New York, NY, USA},
  \bibinfo{pages}{1222–1230}.
\newblock
\showISBNx{9781450321747}
\urldef\tempurl%
\url{https://doi.org/10.1145/2487575.2488200}
\showDOI{\tempurl}


\bibitem[Qu et~al\mbox{.}(2016)]%
        {qu2016}
\bibfield{author}{\bibinfo{person}{Yanru Qu}, \bibinfo{person}{Han Cai},
  \bibinfo{person}{Kan Ren}, \bibinfo{person}{Weinan Zhang},
  \bibinfo{person}{Yong Yu}, \bibinfo{person}{Ying Wen}, {and}
  \bibinfo{person}{Jun Wang}.} \bibinfo{year}{2016}\natexlab{}.
\newblock \showarticletitle{Product-based neural networks for user response
  prediction}. In \bibinfo{booktitle}{\emph{2016 IEEE 16th International
  Conference on Data Mining (ICDM)}}. IEEE, \bibinfo{pages}{1149--1154}.
\newblock


\bibitem[Rendle(2010)]%
        {rendle2010factorization}
\bibfield{author}{\bibinfo{person}{Steffen Rendle}.}
  \bibinfo{year}{2010}\natexlab{}.
\newblock \showarticletitle{Factorization machines}. In
  \bibinfo{booktitle}{\emph{2010 IEEE International Conference on Data
  Mining}}. IEEE, \bibinfo{pages}{995--1000}.
\newblock


\bibitem[Song et~al\mbox{.}(2019)]%
        {song2019autoint}
\bibfield{author}{\bibinfo{person}{Weiping Song}, \bibinfo{person}{Chence Shi},
  \bibinfo{person}{Zhiping Xiao}, \bibinfo{person}{Zhijian Duan},
  \bibinfo{person}{Yewen Xu}, \bibinfo{person}{Ming Zhang}, {and}
  \bibinfo{person}{Jian Tang}.} \bibinfo{year}{2019}\natexlab{}.
\newblock \showarticletitle{Autoint: Automatic feature interaction learning via
  self-attentive neural networks}. In \bibinfo{booktitle}{\emph{Proceedings of
  the 28th ACM International Conference on Information and Knowledge
  Management}}. \bibinfo{pages}{1161--1170}.
\newblock


\bibitem[Wang et~al\mbox{.}(2017)]%
        {wang2017deep}
\bibfield{author}{\bibinfo{person}{Ruoxi Wang}, \bibinfo{person}{Bin Fu},
  \bibinfo{person}{Gang Fu}, {and} \bibinfo{person}{Mingliang Wang}.}
  \bibinfo{year}{2017}\natexlab{}.
\newblock \showarticletitle{Deep \& cross network for ad click predictions}. In
  \bibinfo{booktitle}{\emph{Proceedings of the ADKDD'17}}. ACM,
  \bibinfo{pages}{12}.
\newblock


\bibitem[Wang et~al\mbox{.}(2021)]%
        {WangSCJLHC21}
\bibfield{author}{\bibinfo{person}{Ruoxi Wang}, \bibinfo{person}{Rakesh
  Shivanna}, \bibinfo{person}{Derek~Zhiyuan Cheng}, \bibinfo{person}{Sagar
  Jain}, \bibinfo{person}{Dong Lin}, \bibinfo{person}{Lichan Hong}, {and}
  \bibinfo{person}{Ed~H. Chi}.} \bibinfo{year}{2021}\natexlab{}.
\newblock \showarticletitle{{DCN} {V2:} Improved Deep {\&} Cross Network and
  Practical Lessons for Web-scale Learning to Rank Systems}. In
  \bibinfo{booktitle}{\emph{{WWW} '21: The Web Conference 2021, Virtual Event /
  Ljubljana, Slovenia, April 19-23, 2021}}. \bibinfo{pages}{1785--1797}.
\newblock


\bibitem[Xiao et~al\mbox{.}(2017)]%
        {xiao2017attentional}
\bibfield{author}{\bibinfo{person}{Jun Xiao}, \bibinfo{person}{Hao Ye},
  \bibinfo{person}{Xiangnan He}, \bibinfo{person}{Hanwang Zhang},
  \bibinfo{person}{Fei Wu}, {and} \bibinfo{person}{Tat-Seng Chua}.}
  \bibinfo{year}{2017}\natexlab{}.
\newblock \showarticletitle{Attentional factorization machines: Learning the
  weight of feature interactions via attention networks}.
\newblock \bibinfo{journal}{\emph{arXiv preprint arXiv:1708.04617}}
  (\bibinfo{year}{2017}).
\newblock


\bibitem[Yang et~al\mbox{.}(2020)]%
        {onn}
\bibfield{author}{\bibinfo{person}{Yi Yang}, \bibinfo{person}{Baile Xu},
  \bibinfo{person}{Shaofeng Shen}, \bibinfo{person}{Furao Shen}, {and}
  \bibinfo{person}{Jian Zhao}.} \bibinfo{year}{2020}\natexlab{}.
\newblock \showarticletitle{Operation-aware Neural Networks for user response
  prediction}.
\newblock \bibinfo{journal}{\emph{Neural Networks}}  \bibinfo{volume}{121}
  (\bibinfo{year}{2020}), \bibinfo{pages}{161--168}.
\newblock
\showISSN{0893-6080}
\urldef\tempurl%
\url{https://doi.org/10.1016/j.neunet.2019.09.020}
\showDOI{\tempurl}


\bibitem[Zhang et~al\mbox{.}(2019)]%
        {fat}
\bibfield{author}{\bibinfo{person}{Junlin Zhang}, \bibinfo{person}{Tongwen
  Huang}, {and} \bibinfo{person}{Zhiqi Zhang}.}
  \bibinfo{year}{2019}\natexlab{}.
\newblock \showarticletitle{FAT-DeepFFM: Field Attentive Deep Field-aware
  Factorization Machine}. In \bibinfo{booktitle}{\emph{Industrial Conference on
  Data Mining}}.
\newblock
\urldef\tempurl%
\url{https://api.semanticscholar.org/CorpusID:155099971}
\showURL{%
\tempurl}


\bibitem[Zhu et~al\mbox{.}(2020)]%
        {jieming-2009-05794}
\bibfield{author}{\bibinfo{person}{Jieming Zhu}, \bibinfo{person}{Jinyang Liu},
  \bibinfo{person}{Shuai Yang}, \bibinfo{person}{Qi Zhang}, {and}
  \bibinfo{person}{Xiuqiang He}.} \bibinfo{year}{2020}\natexlab{}.
\newblock \showarticletitle{FuxiCTR: An Open Benchmark for Click-Through Rate
  Prediction}.
\newblock \bibinfo{journal}{\emph{ArXiv}}  \bibinfo{volume}{abs/2009.05794}
  (\bibinfo{year}{2020}).
\newblock


\end{thebibliography}

\end{document}